\documentclass[superscriptaddress,prd,aps,showpacs,nofootinbib,showkeywords,eqsecnum,preprint]{revtex4-1}

\usepackage{graphicx,color,amsmath,amsxtra}
\usepackage{epsf}
\usepackage{amssymb}
\usepackage{enumerate}
\usepackage{hhline}
\usepackage{array}
\usepackage{tabularx}
\usepackage[unicode]{hyperref}
\usepackage{graphicx}                
\usepackage{epstopdf}

\begin{document}
\pagestyle{myheadings}
\title{Stability investigations  of isotropic and anisotropic exponential inflation in the Starobinsky-Bel-Robinson gravity}
\author{Tuan Q. Do}
\email{tuan.doquoc@phenikaa-uni.edu.vn}
\affiliation{Phenikaa Institute for Advanced Study, Phenikaa University, Hanoi 12116, Vietnam}
\affiliation{Faculty of Basic Sciences, Phenikaa University, Hanoi 12116, Vietnam}
\author{Duy H. Nguyen}
\email{duy.nguyenhoang@phenikaa-uni.edu.vn}
\affiliation{Phenikaa Institute for Advanced Study, Phenikaa University, Hanoi 12116, Vietnam}
\author{Tuyen M. Pham}
\email{tuyen.phammanh@phenikaa-uni.edu.vn}
\affiliation{Phenikaa Institute for Advanced Study, Phenikaa University, Hanoi 12116, Vietnam}
\date{\today} 

\begin{abstract}
In this paper, we would like to examine whether a novel Starobinsky-Bel-Robinson gravity model admits stable exponential inflationary solutions with or without spatial anisotropies. As a result, we are able to derive an exact de Sitter inflationary to this Starobinsky-Bel-Robinson model. Furthermore, we observe that an exact Bianchi type I inflationary solution does not exist in the Starobinsky-Bel-Robinson model. However, we find that a modified Starobinsky-Bel-Robinson model, in which the sign of coefficient of $R^2$ term is flipped from positive to negative, can admit the corresponding Bianchi type I inflationary solution. Unfortunately, stability analysis using the dynamical system approach indicates that both of these inflationary solutions turn out to be unstable. Interestingly, we show that a stable de Sitter inflationary solution can be obtained in the modified Starobinsky-Bel-Robinson gravity.
\end{abstract}
\maketitle
\section{Introduction} \label{intro}
Among the very first inflationary models \cite{Starobinsky:1980te,Guth:1980zm,Linde:1981mu,Linde:1983gd}, the Starobinsky model \cite{Starobinsky:1980te} has still remained as one of the most viable models in the light of the Planck observation used to probe the cosmic microwave background radiations (CMB) \cite{Planck}. The success of Starobinsky model is due to the inclusion of the Ricci scalar squared term $R^2$, which acts as a quantum correction \cite{Starobinsky:1980te}. Interestingly, one can derive the corresponding effective action of scalar field from the Starobinsky model by using a suitable conformal transformation \cite{Whitt:1984pd,Maeda:1987xf,Barrow:1988xh} and can therefore derive its theoretical predictions for the CMB probes, e.g., see Refs. \cite{Sebastiani:2013eqa,Mishra:2018dtg,Mishra:2019ymr,Shtanov:2022pdx}. 

It appears that the Starobinsky model is the simplest higher-order extension of the Einstein's gravity \cite{Nojiri:2010wj,Nojiri:2017ncd}. In high energy physics, gravity models having higher-order curvature terms have played a leading candidate for searching an ultraviolet (UV) completeness of Einstein's general relativity \cite{Koshelev:2017tvv}. This is based on the fact that the pure Einstein's gravity is non-renormalizable and therefore cannot be quantized.  However, by adding higher-order curvature correction terms such as the quadratic ones, $R^2$ and $R_{\mu\nu}R^{\mu\nu}$, into the pure Einstein-Hilbert action, Stelle has been able to obtain a renormalizable model \cite{Stelle:1976gc}. 
It should be noted that the existence of quadratic curvature terms will lead to the appearance of fourth-order derivatives in Einstein field equations. Therefore, quadratic gravity models are well known as fourth-order gravity ones \cite{Starobinsky:1987zz}, whose rich history along with physical and cosmological aspects have been summarized in interesting reviews \cite{Schmidt:2006jt,Salvio:2018crh}. Beside inflationary universes, fourth-order gravities have been regarded as a promising approach to reveal the nature of the cosmic acceleration, e.g., see Refs. \cite{Nojiri:2010wj,Nojiri:2017ncd,Carroll:2004de} for details. 

 It should be noted that the existence of higher-than-two derivatives could destroy the stability of the quadratic gravity due to the associated  Ostrogradsky ghosts \cite{Woodard:2015zca}. Very interestingly, however, the Starobinsky model seems to be the special quadratic gravity model being free of this ghost as indicated in Ref. \cite{Woodard:2015zca}. This result together with the success in predicting an inflationary universe make the Starobinsky model very unique and attractive.  However, the precise CMB measurements may address refinements of the predictions of the Starobinsky model. Indeed, various extensions of the Starobinsky model have been proposed recently, e.g., see Refs. \cite{Appleby:2009uf,Myrzakulov:2014hca,Netto:2015cba,Myrzakulov:2016tsz,Elizalde:2017mrn,Liu:2018hno,Aldabergenov:2018qhs,Elizalde:2018now,Elizalde:2018rmz,Cano:2020oaa,Rodrigues-da-Silva:2021jab,Ivanov:2021chn,Koshelev:2022olc,Modak:2022gol,Ketov:2022lhx,CamposDelgado:2022sgc,Ketov:2022zhp}. Among them, we are currently interested in the so-called Starobinsky-Bel-Robinson (SBR) gravity  proposed by Ketov \cite{Ketov:2022lhx}, whose action involves not only the $R^2$ term but also the known Bel-Robinson (BR) tensor squared \cite{Bel:1959uwe,Robinson:1959ev,Deser:1999jw}, which is quartic in the curvature and can be interpreted  as a superstring-inspired quantum correction \cite{Iihoshi:2007vv}. In the follow-up papers, Ketov and his colleagues have found the corresponding Schwarzschild-type black holes \cite{CamposDelgado:2022sgc} as well as isotropic inflationary solutions \cite{Ketov:2022zhp} to this novel fourth-order gravity model. Moreover, they have pointed out that the de Sitter inflationary solution of the SBR model is unstable against perturbations, while a perturbative solution of the SBR model turns out to be an attractor one. It turns out that investigating whether a gravity model admits either stable or unstable de Sitter solutions is an important task because it would tell us which phase of our universe the model would be suitable for \cite{Elizalde:2014xva,Pozdeeva:2019agu}. See also Refs. \cite{Appleby:2009uf,Netto:2015cba} for related discussions on the (in)stability of de Sitter solutions of some non-trivial extensions of the Starobinsky model.
  
 Motivated by our past experiences of showing a no-go theorem for exponential inflation in the so-called Ricci-inverse gravity \cite{Do:2020vdc}, which is another  novel fourth-order gravity model proposed by Amendola and his colleagues \cite{Amendola:2020qho}, we really want to examine whether the SBR gravity model admits stable (an)isotropic exponential inflationary solutions. It is noted that if the SBR gravity model admitted a stable anisotropic inflation it would be a counterexample to the so-called cosmic no-hair conjecture proposed by Hawking and his colleagues long time ago \cite{GH}. Basically, this conjecture states that the late time universe would obey the cosmological principle, i.e., would be homogeneous and isotropic on large scales, regardless of its early states. Many people have made huge efforts to prove this conjecture since the seminal paper of Wald for Bianchi spacetimes in the presence of cosmological constant $\Lambda$ \cite{Wald:1983ky,Barrow:1987ia,Mijic:1987bq,Kitada:1991ih,Maleknejad:2012as}, but a general proof has remained unknown until now. In cosmology, Bianchi spacetimes are homogeneous but anisotropic metrics \cite{bianchi}. Very interestingly, Starobinsky showed in his seminal paper \cite{Starobinsky:1982mr} that the cosmic no-hair conjecture should be valid locally, i.e., inside of the future event horizon,  if it is correct. This result was then confirmed by other works \cite{Muller:1989rp,Barrow:1984zz}. Beside the proofs, counterexamples to the cosmic no-hair conjecture have been claimed to exist in many gravity models. An interesting example can be found in papers written by Barrow and Hervik, in which they suggested that stable Bianchi inflationary solutions could emerge within quadratic gravities \cite{barrow05,barrow06}. In a follow-up paper, Middleton has arrived at the similar conclusion for higher-order theories of gravity \cite{Middleton:2010bv}. However, Kao and Lin have shown in their papers \cite{kao09} that some Bianchi solutions found in Refs. \cite{barrow05,barrow06} turn out to be unstable against field perturbations. This means that the validity of the cosmic no-hair conjecture may not be violated within these  quadratic gravity models. Other important stability analysis of Bianchi type I inflationary solutions within not only the Starobinsky but also quadratic gravity models can be found in Refs. \cite{Toporensky:2006kc,Muller:2017nxg}. It is therefore important to check if any extensions or modifications of the Starobinsky gravity model admit counterexamples to the cosmic no-hair conjecture. 
 
 In summary, this paper will be organized as follows: (i) Its brief introduction has been written in Sec. \ref{intro}. (ii) Basic setup of the Starobinsky-Bel-Robinson gravity model along with its (an)isotropic exponential inflationary solutions will be presented in Sec. \ref{sec2}.  (iii) Stability of the obtained inflationary solutions will be analyzed via the dynamical system method in Sec. \ref{sec3}. (iv) Finally, concluding remarks will be given in Sec. \ref{final}. Some additional useful calculations will be listed in the Appendices. 
\section{Basic setup} \label{sec2}
\subsection{The Starobinsky-Bel-Robinson model}
The action of the Starobinsky-Bel-Robinson (SBR) gravity has been proposed by Ketov as follows \cite{Ketov:2022lhx,CamposDelgado:2022sgc,Ketov:2022zhp} 
\begin{equation}
S_{\rm SBR}= \frac{M_p^2}{2}\int d^4 x \sqrt{-g}  \left[R+\frac{1}{6m^2} R^2 -\frac{\beta}{8m^6}T^{\mu\nu\lambda\rho}T_{\mu\nu\lambda\rho} \right],
\end{equation}
where $T^{\mu\nu\lambda\rho}$ is the Bel-Robinson (BR) tensor in four-dimensional spacetimes, whose definition is given by \cite{Bel:1959uwe,Robinson:1959ev,Deser:1999jw}
\begin{equation}
T^{\mu\nu\lambda\rho} =R^{\mu\alpha\beta\lambda} R^\nu{}_{\alpha\beta}{}^\rho +R^{\mu\alpha\beta\rho}R^\nu{}_{\alpha\beta}{}^\lambda-\frac{1}{2}g^{\mu\nu}R^{\alpha\beta\gamma\lambda}R_{\alpha\beta\gamma}{}^\rho.
\end{equation}
For convenience, the above action has been rewritten in a more transparent form such as 
\begin{equation} 
S_{\rm SBR}= \frac{M_p^2}{2}\int d^4 x \sqrt{-g}  \left[R+\alpha_1 R^2 +\alpha_2 \left(E_4^2- P_4^2\right)\right],
\end{equation}
due to the identity derived in Ref. \cite{Deser:1999jw},
\begin{equation}
T^{\mu\nu\lambda\rho}T_{\mu\nu\lambda\rho}  =\frac{1}{4} \left(P_4^2-E_4^2 \right),
\end{equation}
where $E_4$ and $P_4$ are the Euler and Pontryagin (topological) densities in four dimensions \cite{Ketov:2022lhx,CamposDelgado:2022sgc,Ketov:2022zhp}. It is very interesting that the Euler density $E_4$ has been shown to be identical to the Gauss-Bonnet (GB) term ${\cal G}$ and the action  $S_{\rm SBR}$ therefore becomes as \cite{Ketov:2022lhx,CamposDelgado:2022sgc,Ketov:2022zhp}
\begin{equation} \label{action}
S_{\rm SBR}= \frac{M_p^2}{2}\int d^4 x \sqrt{-g}  \left[R+\alpha_1 R^2 +\alpha_2 \left({\cal G}^2-{P_4}^2\right)\right],
\end{equation}
here the definition of ${\cal G}$ and $P_4$ are given by 
\begin{align}
{\cal G} &= R^2 -4R_{\mu\nu} R^{\mu\nu} +R_{\mu\nu\rho\sigma}R^{\mu\nu\rho\sigma},\\
{P_4}&= \frac{1}{2}\sqrt{-g} \epsilon_{\mu\nu\rho\sigma} R^{\rho \sigma}{}_{\alpha \beta} R^{\mu\nu\alpha\beta},
\end{align}
where $\epsilon_{\mu\nu\rho\sigma}$ is the totally antisymmetric Levi-Civita tensor with $\epsilon_{0123}=1$. In the above action, we have introduced new parameters as 
\begin{equation}
\alpha_1 =\frac{1}{6m^2}, ~ \alpha_2 =\frac{ \beta}{32m^6},
\end{equation}
 for convenience. Additionally, $M_p$ is the reduced Planck mass. In the pure Starobinsky gravity \cite{Starobinsky:1980te}, $m$ is nothing but the parameter determining the mass of inflaton field (a.k.a. the scalaron mass). In the SBR gravity, $\beta$ has been introduced as a positive dimensionless coupling constant \cite{Ketov:2022zhp}, whose (unknown) value is supposed to be determined by compactification of $M$-theory \cite{CamposDelgado:2022sgc}.  It is clear that both ${\cal G}$ and ${P_4}$ are terms quartic in the curvature. The corresponding tensorial field equations of the SBR gravity have been derived in Refs. \cite{CamposDelgado:2022sgc,Ketov:2022zhp}. In a case of vanishing ${ P_4}$, this SBR gravity can be regarded as a specific scenario of the $F(R,{\cal G})$ \cite{DeLaurentis:2015fea}. Interestingly, the cosmological inflation has been investigated in the model with $F(R,{\cal G})=R+\alpha_1 R^2 +\alpha_2 {\cal G}^2$ \cite{DeLaurentis:2015fea} as well as in a recent paper on the SBR gravity model \cite{Ketov:2022zhp}. 
\subsection{Isotropic inflation}
Now, we would like to see whether the SBR gravity admits a homogeneous and  isotropic spacetime as its cosmological solutions. We therefore consider the following Friedmann-Lemaitre-Robertson-Walker (FLRW) metric,
\begin{equation} \label{metric}
ds^2 =-N^2(t)dt^2 +e^{2\alpha(t)} \left(dx^2 +dy^2 +dz^2 \right),
\end{equation}
where  $N(t)$ is the lapse function introduced to obtain the following Friedmann equations from its Euler-Lagrange equation \cite{Myrzakulov:2014hca,Do:2020vdc,Toporensky:2006kc,Kao:1991zz}, while $\alpha(t)$ is a scale factor. It is noted that $N$ should be set to be one after deriving its corresponding Friedmann equation along with other field equations of scale factors as done in many previous papers, e.g., see Refs. \cite{Myrzakulov:2014hca,Do:2020vdc,Toporensky:2006kc,Kao:1991zz}.   As a result, the corresponding Gauss-Bonnet density ${\cal G}$ is defined to be
\begin{align}
{\cal G} =-\frac{24}{N^5} \dot\alpha^2 \left[ \dot N \dot \alpha - N \left(\ddot\alpha+\dot\alpha^2 \right) \right],
\end{align}
while the Pontryagin density ${P_4}$ vanishes for the FLRW metric  due to its spherical symmetry \cite{Ketov:2022zhp}. It is noted that ${ P_4}$ also vanishes for a Schwarzschild-type black hole metric as mentioned in Ref. \cite{CamposDelgado:2022sgc}. In order to derive the corresponding differential  field equations of the SBR gravity for this FLRW metric, we will consider the effective calculation approach based on the Euler-Lagrange equations, which have been widely used not only in our previous papers \cite{Do:2020vdc} but also in other papers  written by other people, e.g., see Ref. \cite{Myrzakulov:2014hca}. Ones can, of course,  derive the corresponding differential field equations for the Bianchi type I metric using the tensorial field equations derived in the original papers of the SBR gravity \cite{Ketov:2022lhx,CamposDelgado:2022sgc,Ketov:2022zhp}. 

As a first step, we will define an explicit expression of the Lagrangian of the SBR model, which is given by
 \begin{equation}
 {\cal L} =\sqrt{-g} \left(R+\alpha_1 R^2 +\alpha_2 {\cal G}^2 \right).
 \end{equation}
 It turns out that $\sqrt{-g}=N e^{3\alpha}$ and $R= -6N^{-2} \left(N^{-1} \dot N \dot\alpha -\ddot\alpha-2\dot\alpha^2 \right) $ for the FLRW metric. Here, it is understood that $\dot\alpha \equiv d\alpha/dt$ as well as $\ddot\alpha \equiv d^2\alpha/dt^2$.  It appears that ${\cal L}$ contains the second-order derivative in time of $\alpha$ and the first-order derivative in time of $N$. Therefore, the corresponding Euler-Lagrange equations for $N$ and $\alpha$ are given by
\begin{align}
\frac{\partial {\cal L}}{\partial N} -\frac{d}{dt} \left(\frac{\partial {\cal L}}{\partial \dot N}\right)&=0,\\
\frac{\partial {\cal L}}{\partial \alpha} -\frac{d}{dt} \left(\frac{\partial {\cal L}}{\partial \dot \alpha}\right) + \frac{d^2}{dt^2} \left(\frac{\partial {\cal L}}{\partial \ddot\alpha}\right)&=0,
\end{align}
respectively.  As a result, these Euler-Lagrange equations for the FLRW metric become
\begin{align} \label{equation-1-FLRW}
&6\alpha_1 \left( 2 \dot\alpha \alpha^{(3)} -\ddot\alpha^2 +6 \dot\alpha^2 \ddot\alpha \right) + 96 \alpha_2 \dot\alpha^4 \left(2 \dot\alpha \alpha^{(3)} +3 \ddot\alpha^2 +6\dot\alpha^2 \ddot\alpha - \dot\alpha^4 \right) + \dot\alpha^2 =0,\\
\label{equation-2-FLRW}
&6 \alpha_1 \left( 2 \alpha^{(4)} + 12\dot\alpha \alpha^{(3)} +9 \ddot\alpha^2 +18 \dot\alpha \ddot\alpha \right)  +96 \alpha_2  \dot\alpha^2 \left[ 2 \dot\alpha^2 \alpha^{(4)}  + \dot\alpha  \left( 16 \ddot\alpha + 12 \dot\alpha^2 \right) \alpha^{(3)} +12 \ddot\alpha^3  \right. \nonumber\\
&\left.+45 \dot\alpha^2 \ddot\alpha^2+10 \dot\alpha^4 \ddot\alpha - 3 \dot\alpha^6  \right] +2 \ddot\alpha +3 \dot\alpha^2 =0,
\end{align}
respectively. Here, $\alpha^{(n)} \equiv d^n \alpha/dt^n$ as a $n$th-order derivative. 
It appears that these field equations coincide with that derived in Ref. \cite{DeLaurentis:2015fea} as well as with that derived in Ref. \cite{Ketov:2022zhp} if we set $a(t)=e^{\alpha(t)}$, or equivalently $H(t)\equiv \dot a/a = \dot\alpha$. 

Now, we would like to figure out an exact de Sitter solution with the scale factor being an exponential function of cosmic time $t$ to these field equations, similar to the previous investigations on fourth-order gravities \cite{Do:2020vdc,barrow05,barrow06,Toporensky:2006kc}, by taking an ansatz for the scale factor 
\begin{equation}
\alpha =\zeta t.
\end{equation}
It is well-known that the pure Starobinsky model with vanishing $\alpha_2$ does not admit an exact de Sitter inflationary solution \cite{Ketov:2022zhp}.
As a result, both of two field equations, i.e., Eqs. \eqref{equation-1-FLRW} and \eqref{equation-2-FLRW}, lead to the same equation of $\zeta$,
\begin{equation} \label{equation-of-zeta-iso}
96 \alpha_2  \zeta^6 -1 =0,
\end{equation}
which gives us a non-trivial exact solution of $\zeta$,
\begin{equation} \label{sol-of-zeta-FLRW}
\zeta =\left( \frac{1}{96\alpha_2} \right)^{\frac{1}{6}}.
\end{equation}
It is now clear that $0<\alpha_2 \ll 1$ is an efficient constraint in order to have the inflation with $\zeta \gg 1$. The positivity constraint of $\alpha_2$ is indeed consistent with the requirement $\beta >0$. This de Sitter solution is exactly that found in a recent paper \cite{Ketov:2022zhp}. It is clear that $\alpha_1$ contributes nothing to this solution. 

It appears, according to Eq. \eqref{sol-of-zeta-FLRW}, that the scale factor is solely determined by $\alpha_2$ rather than $\alpha_1$. This result is consistent with investigations made in Refs. \cite{barrow05,Toporensky:2006kc}, in which the quadratic curvature terms contribute nothing to the de Sitter solution. However, the existence of the cosmological constant $\Lambda$ ensures the appearance of  such inflation \cite{barrow05,Toporensky:2006kc}. In the present paper, the BR term can be regarded as an effective potential causing the corresponding inflationary phase. 

It should be noted that the authors of paper \cite{DeLaurentis:2015fea} have considered two regimes to figure out approximated solutions to Eqs. \eqref{equation-1-FLRW} and \eqref{equation-2-FLRW}. The first one associated with the case that $6\alpha_1 \gg 96\alpha_2 \dot\alpha^4$ will yield the following solution $\dot\alpha^2 =1/(6\alpha_1)$ with a scalaron mass $m_R^2 \equiv 1/(6 \alpha_1)$. The second one associated with the case that $96\alpha_2 \dot\alpha^4 \gg 6\alpha_1$ will imply the corresponding solution $\dot\alpha^6 = 1/(96\alpha_2)$ with another scalaron mass $m_{\cal G}^2 \equiv 1/\sqrt[3]{96\alpha_2}$. It is now clear that the exact solution derived above is similar to that found in the regime $96\alpha_2 \dot\alpha^4 \gg 6\alpha_1$. 

It is important to note that the investigation in Ref. \cite{Ketov:2022zhp} has  verified that the exact de Sitter solution shown above turns out to be unattractive and unstable against perturbations of $H_{\rm dS}$. Instead, a perturbative solution of the SBR model, in which a slow-roll approximate solution found in the pure Starobinsky gravity model plays a leading role while the BR term is nothing but a superstring-inspired quantum correction, has been figured out, regarding the parameter $\alpha_2$ to be small \cite{Ketov:2022zhp}. Interestingly, this perturbative solution has been shown to be attractive.
\subsection{Anisotropic inflation}
In this subsection, we extend our analysis to the Bianchi type I metric, which is nothing but a homogeneous and anisotropic spacetime \cite{Do:2020vdc},
\begin{equation} \label{BI-metric}
ds^2 =-N^2(t)dt^2 +e^{2\alpha(t)-4\sigma(t)}dx^2 +e^{2\alpha(t)+2\sigma(t)} \left(dy^2 +dz^2 \right),
\end{equation}
where $\sigma(t)$ acts as a deviation from isotropy and therefore it should be much smaller than $\alpha(t)$. It is clear that the isotropic spacetime corresponds to $\sigma(t)=0$. It should be noted that this ansatz is a special case of that used in Refs. \cite{barrow05,barrow06,Muller:2017nxg} with $\sigma_+=\sigma$  and $\sigma_- =0$. The reason is simply that we would like to figure out exact analytical solutions to the SBR model, similar to what we have done in different models \cite{Do:2020vdc,Kanno:2010nr,Do:2011zza}.  To be more specific, we will briefly demonstrate in the following subsection \ref{general case} that the inclusion of non-vanishing $\sigma_-$ will make the number of field variables greater than the number of independent field equations. Therefore, it is difficult to figure out exact analytical solutions of $\alpha$, $\sigma_+$, as well as $\sigma_-$ from the field equations in the case of non-vanishing $\sigma_-$. Of course, non-trivial relations among these three scale factors can be figured out. It turns out that the existence of non-vanishing $\sigma_-$ insignificantly affects  the value of $\alpha$ and $\sigma_+$ during the inflationary phase as expected.

Additionally, although the Bianchi type I metric breaks the spherical symmetry, the corresponding Pontryagin density ${P_4}$ still contributes nothing to the dynamics of spacetime due to the fact that all Riemann tensors with completely different indices, i.e., $\mu=0$, $\nu=1$, $\alpha=2$, and $\beta =3$, vanish automatically for the Bianchi type I metric. As a result, the corresponding Ricci scalar and Gauss-Bonnet term are given by
\begin{align}
R & =-\frac{6}{N^2} \left( \frac{\dot N}{N} \dot\alpha   -\ddot\alpha - 2\dot\alpha^2-\dot\sigma^2 \right),\\
{\cal G} &=-\frac{24}{N^5}\left( \dot\alpha +\dot\sigma \right) \left\{ \dot N \left(\dot \alpha -2\dot\sigma \right)\left(\dot\alpha+\dot\sigma\right) - N \left[\ddot\alpha \left(\dot\alpha-\dot\sigma \right)-2\ddot\sigma \dot\sigma +\dot\alpha^3 -\dot\alpha\dot\sigma \left(\dot\alpha+2\dot\sigma \right)   \right] \right\}.
\end{align}
Thanks to these results,  we are able to define the corresponding Euler-Lagrange equations. As a result, the Euler-Lagrange equation for $N$ is given by
\begin{equation}
\frac{\partial {\cal L}}{\partial N} -\frac{d}{dt} \left(\frac{\partial {\cal L}}{\partial \dot N}\right)=0,
\end{equation}
which can be defined explicitly to be a third-order ordinary differential equation (ODE),
\begin{align} \label{N-field-eq-BI}
&6\alpha_1 \left[ 2\dot\alpha \alpha^{(3)} - \ddot\alpha \left(\ddot\alpha -6\dot\alpha^2 +4\dot\sigma^2 \right)+ 4\dot\alpha \dot\sigma \ddot\sigma - 3 \dot\sigma^2 \left(2\dot\alpha^2 +\dot\sigma^2 \right) \right] \nonumber\\
&+96\alpha_2 \left(\dot\alpha+\dot\sigma \right)^2 \left[2 \alpha^{(3)} \left(\dot\alpha-2\dot\sigma \right)\left(\dot\alpha^2-\dot\sigma^2 \right) -4\sigma^{(3)} \dot\sigma \left(\dot\alpha+\dot\sigma\right) \left(\dot\alpha-2\dot\sigma \right)  +\ddot\alpha^2 \left( 3\dot\alpha^2  -6\dot\alpha \dot\sigma-\dot\sigma^2 \right) \right. \nonumber\\
&\left. -4\ddot\sigma^2 \left(\dot\alpha^2 -3\dot\sigma^2 \right) -4\ddot\alpha \ddot\sigma \dot\sigma \left( \dot\alpha  -3\dot\sigma \right)  +2\ddot\alpha \left(\dot\alpha+\dot\sigma \right) \left(\dot\alpha-2\dot\sigma\right) \left(3\dot\alpha^2-3\dot\alpha \dot\sigma -2\dot\sigma^2 \right) \right. \nonumber\\
&\left. -8\ddot\sigma \dot\alpha \dot\sigma \left(\dot\alpha+\dot\sigma\right)  \left(\dot\alpha  -2  \dot\sigma \right)  -\dot\alpha^2\left(\dot\alpha+\dot\sigma \right)^2 \left(\dot\alpha-2\dot\sigma \right)^2 \right] +\dot\alpha^2-\dot\sigma^2  =0.
\end{align}
 This is nothing but the Friedmann equation. On the other hand, the Euler-Lagrange equation for $\alpha$ turns out to be
\begin{equation}
\frac{\partial {\cal L}}{\partial \alpha} -\frac{d}{dt} \left(\frac{\partial {\cal L}}{\partial \dot \alpha}\right) + \frac{d^2}{dt^2} \left(\frac{\partial {\cal L}}{\partial \ddot\alpha}\right)=0,
\end{equation}
which can be defined to be a fourth-order ODE,
\begin{align}  \label{alpha-field-eq-BI}
&6\alpha_1 \left[ 2\alpha^{(4)} +12\dot\alpha \alpha^{(3)} +4\dot\sigma \sigma^{(3)} + 9\ddot\alpha^2 +4\ddot\sigma^2 +2\ddot\alpha \left(9\dot\alpha^2 +2\dot\sigma^2 \right) +8\ddot\sigma \dot\alpha\dot\sigma  +3\dot\sigma^2 \left(2\dot\alpha^2+\dot\sigma^2\right) \right] \nonumber\\
& +96\alpha_2 \biggl\{ 2\alpha^{(4)} \left(\dot\alpha^2 -\dot\sigma^2  \right)^2 -4\sigma^{(4)} \dot\sigma \left(\dot\alpha - \dot\sigma\right) \left(\dot\alpha +\dot\sigma \right)^2   \nonumber\\
&  +4\alpha^{(3)} \left(\dot\alpha^2-\dot\sigma^2 \right) \left[ 4 \ddot\alpha \dot\alpha -4\ddot\sigma \dot\sigma +3\dot\alpha \left(\dot\alpha^2 -\dot\sigma^2 \right) \right]  \nonumber\\
& - 4\sigma^{(3)} \left(\dot\alpha+\dot\sigma \right) \left[ \ddot\alpha \dot\sigma \left(5\dot\alpha -3\dot\sigma \right)+ \ddot\sigma \left(3\dot\alpha^2 +3\dot\alpha \dot\sigma -8\dot\sigma^2 \right) + \dot\sigma \left(\dot\alpha+\dot\sigma \right)  \left(5\dot\alpha^2-5\dot\alpha \dot\sigma +2\dot\sigma^2 \right) \right] \nonumber\\
&+4\ddot\alpha^3 \left( 3 \dot\alpha^2 - \dot\sigma^2\right) -8\ddot\sigma^3 \left( \dot\alpha^2 -\dot\alpha\dot\sigma -3\dot\sigma^2 \right) -\ddot\alpha^2 \left(24 \dot\alpha \dot\sigma \ddot\sigma -45\dot\alpha^4 -9 \dot\sigma^4+ 54\dot\alpha^2 \dot\sigma^2  \right)  \nonumber\\
& -4 \ddot\sigma^2\left[ \ddot\alpha \left(5\dot\alpha^2 +4\dot\alpha \dot\sigma -7\dot\sigma^2  \right) +5\dot\alpha^4+7\dot\sigma^4 +2 \dot\alpha \dot\sigma \left( \dot\alpha-\dot\sigma  \right) \left(5\dot\alpha +2\dot\sigma\right) \right] \nonumber\\
& -4 \ddot\alpha \ddot\sigma \dot\sigma  \left(\dot\alpha+\dot\sigma\right)  \left(23\dot\alpha^2-14\dot\alpha \dot\sigma -\dot\sigma^2\right) \nonumber\\
&  + 2 \ddot\alpha \left(\dot\alpha+\dot\sigma\right)^2 \left(5\dot\alpha^4 -10\dot\alpha^3 \dot\sigma +15\dot\alpha^2 \dot\sigma^2 -10 \dot\alpha \dot\sigma^3 -4\dot\sigma^4 \right)  \nonumber\\
& - 12 \ddot\sigma \dot\alpha \dot\sigma  \left(\dot\alpha+\dot\sigma \right)^2 \left(\dot\alpha^2 -\dot\alpha \dot\sigma +4\dot\sigma^2 \right)   -3 \dot\alpha^2 \left( \dot\alpha+\dot\sigma  \right)^4 \left(\dot\alpha-2\dot\sigma \right)^2  \biggr\} +2\ddot\alpha+3\dot\alpha^2 +3\dot\sigma^2 =0.
\end{align}
Finally, the  Euler-Lagrange equation for $\sigma$ reads
\begin{equation}
\frac{\partial {\cal L}}{\partial \sigma} -\frac{d}{dt} \left(\frac{\partial {\cal L}}{\partial \dot \sigma}\right) + \frac{d^2}{dt^2} \left(\frac{\partial {\cal L}}{\partial \ddot\sigma}\right)=0,
\end{equation}
whose explicit form is defined to be
\begin{align}  \label{sigma-field-eq-BI}
&12 \alpha_1 \left[ \dot\sigma \alpha^{(3)} +\ddot\alpha \left(\ddot\sigma +7\dot\alpha\dot\sigma \right) + \ddot\sigma \left(2\dot\alpha^2 +3\dot\sigma^2\right) +3\dot\alpha \dot\sigma \left(2\dot\alpha^2  + \dot\sigma^2\right) \right] \nonumber\\
& +192 \alpha_2 \biggl\{ \alpha^{(4)} \dot\sigma \left( \dot\alpha-\dot\sigma \right)\left(\dot\alpha+\dot\sigma\right)^2 -2\sigma^{(4)}\dot\sigma^2 \left(\dot\alpha+\dot\sigma \right)^2  \nonumber\\
& +\alpha^{(3)} \left( \dot\alpha+\dot\sigma \right)  \left[\ddot\alpha  \dot\sigma \left(7\dot\alpha-\dot\sigma \right)+\ddot\sigma   \left(\dot\alpha^2 +\dot\alpha\dot\sigma -8\dot\sigma^2  \right) + \dot\sigma \left(\dot\alpha+\dot\sigma \right) \left(7\dot\alpha^2 -7\dot\alpha\dot\sigma -2\dot\sigma^2 \right) \right]  \nonumber\\
& -4 \sigma^{(3)} \dot\sigma\left(\dot\alpha +\dot\sigma \right)  \left[ 2 \ddot\alpha \dot\sigma +2 \ddot\sigma \left(\dot\alpha+2\dot\sigma \right) +3 \dot\alpha \dot\sigma \left(\dot\alpha+\dot\sigma \right) \right]  \nonumber\\
&+2 \ddot\alpha^3 \dot\sigma \left(2\dot\alpha + \dot\sigma \right) -2\ddot\sigma^3 \left( \dot\alpha^2  +6 \dot\alpha \dot\sigma+6 \dot\sigma^2 \right)  \nonumber\\
& +2\ddot\alpha^2 \left[ \ddot\sigma \left(\dot\alpha^2 +2\dot\alpha \dot\sigma -2\dot\sigma^2 \right) +\dot\sigma \left(\dot\alpha+\dot\sigma \right) \left(11\dot\alpha^2-2\dot\alpha \dot\sigma -4\dot\sigma^2 \right) \right]  \nonumber\\
& -6 \ddot\sigma^2 \dot\sigma \left[\ddot\alpha \left(2\dot\alpha  + 3\dot\sigma \right) +3 \dot\alpha \left(\dot\alpha+\dot\sigma \right) \left( \dot\alpha+ 2  \dot\sigma  \right) \right]  \nonumber\\
& + 4\ddot\alpha \ddot\sigma \left(\dot\alpha+\dot\sigma  \right) \left(\dot\alpha+4\dot\sigma \right) \left(\dot\alpha^2 -3\dot\alpha \dot\sigma -\dot\sigma^2 \right)  \nonumber\\
&  +6 \ddot\alpha \dot\alpha\dot\sigma \left(\dot\alpha+\dot\sigma  \right)^2   \left(2\dot\alpha^2 -2\dot\alpha\dot\sigma -\dot\sigma^2 \right)  -18 \ddot\sigma \dot\alpha^2 \dot\sigma^2 \left(  \dot\alpha  + \dot\sigma\right)^2  \biggr\} +\ddot\sigma +3 \dot\alpha \dot\sigma =0,
\end{align}
which is also a fourth-order ODE. It is straightforward to check that Eqs. \eqref{N-field-eq-BI}, \eqref{alpha-field-eq-BI}, and \eqref{sigma-field-eq-BI} all will recover Eqs. \eqref{equation-1-FLRW} and \eqref{equation-2-FLRW} in the isotropic limit, $\sigma \to 0$.

Now, we are going to find out exact solutions to these field equations by using an ansatz for the scale factors \cite{Do:2020vdc}
 \begin{equation}
\alpha =\zeta t, ~\sigma = \eta t.
\end{equation}
As a result, Eqs. \eqref{N-field-eq-BI}, \eqref{alpha-field-eq-BI},  and \eqref{sigma-field-eq-BI} can be reduced to the corresponding algebraic equations of $\zeta$ and $\eta$,
\begin{align} \label{algebraic-equation-1}
18\alpha_1 \eta^2 \left(2\zeta^2 +\eta^2 \right) +96\alpha_2 \zeta^2 \left(\zeta-2\eta\right)^2 \left(\zeta+\eta   \right)^4 -\zeta^2 +\eta^2 &=0,\\
\label{algebraic-equation-2}
6\alpha_1 \eta^2 \left(2\zeta^2 +\eta^2 \right) -96\alpha_2 \zeta^2 \left(\zeta-2\eta\right)^2 \left(\zeta+\eta   \right)^4 +\zeta^2 +\eta^2 &=0,\\
\label{algebraic-equation-3}
\zeta \eta \left[ 12\alpha_1 \left(2\zeta^2 +\eta^2 \right) +1 \right] &=0,
\end{align}
respectively. The last equation implies a non-trivial one
\begin{equation} \label{reduced-equation-1}
2\zeta^2 +\eta^2 = -\frac{1}{12\alpha_1},
\end{equation}
here we have ignored an isotropic solution $\eta=0$.
Thanks to this relation, the first two equations both reduce to the same equation,
\begin{equation} \label{reduced-equation-2}
192 \alpha_2 \zeta^2  \left(\zeta-2\eta \right)^2 \left(\zeta+\eta \right)^4 +\frac{1}{12\alpha_1} =0.
\end{equation}
It should be noted that $\alpha_1$ has to be positive definite due to its definition, $\alpha_1 \equiv (6m^2)^{-1}$. Hence, Eq. \eqref{reduced-equation-1} cannot admit any real solution of $\zeta$ or $\eta$. This implies that the SBR gravity model cannot admit the Bianchi type I solution for the positive $\alpha_1$. However, if we flip the sign of $\alpha_1$ from positive to negative, i.e., $\alpha_1 \to \alpha_1 = -(6m^2)^{-1}<0$, the corresponding SBR gravity model would admit anisotropic Bianchi type I solutions.

\begin{figure}[hbtp] 
\begin{center}
{\includegraphics[height=50mm]{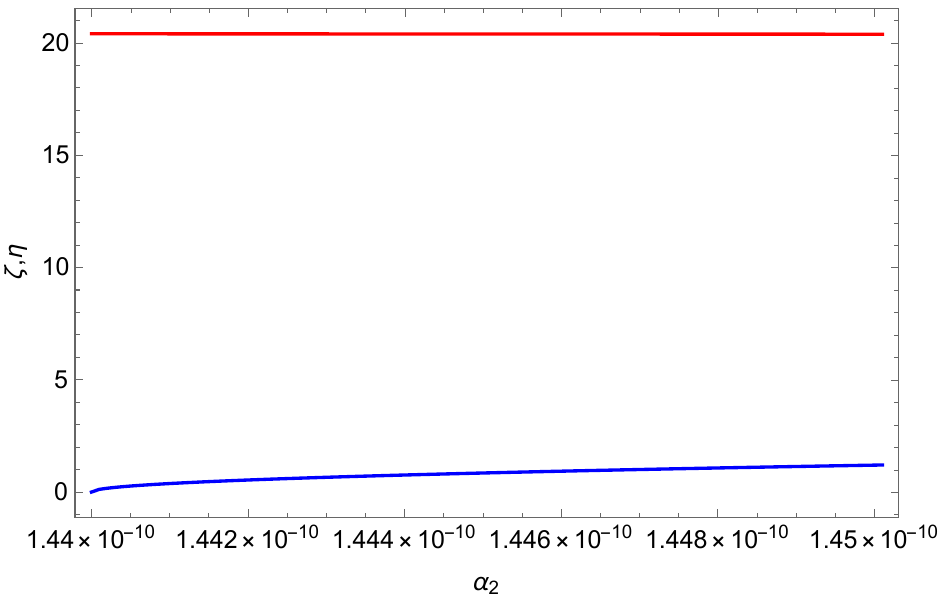}}\quad
{\includegraphics[height=50mm]{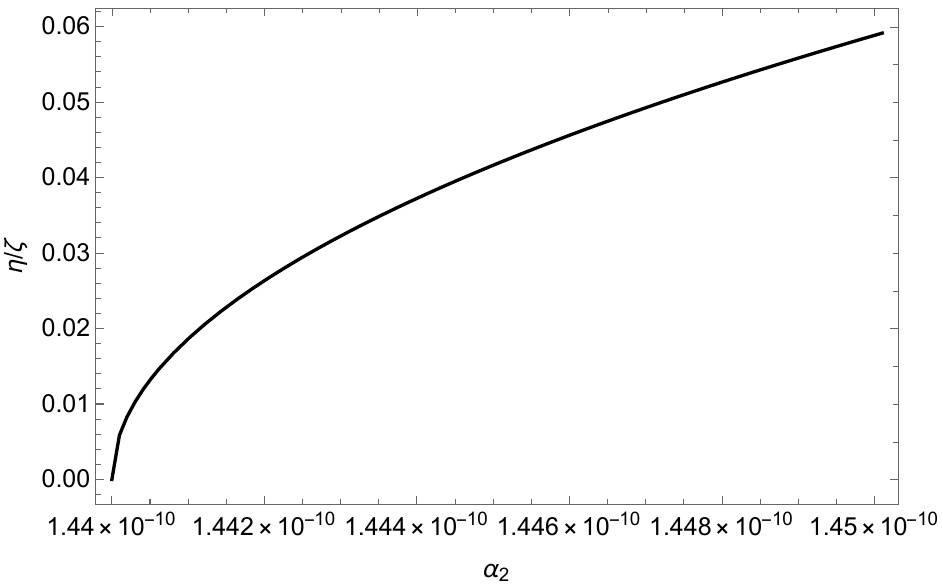}}\\
\caption{(Left) The red and blue curves correspond to the plot of $\zeta$ and $\eta$, respectively, provided that $\alpha_1 =-10^{-4}$.  (Right) The corresponding ratio of $\eta/\zeta$ as a function of $\alpha_2$.}
\label{fig0}
\end{center}
\end{figure}

It appears that $\zeta \gg \eta$ for any viable anisotropic inflationary solutions. Hence,  we can approximately evaluate the value of $\zeta$ as well as $\eta$ as follows, assuming $\alpha_1$ to be negative definite. First, Eq. \eqref{reduced-equation-2} implies that
\begin{equation}
\zeta^8 \simeq -\frac{1}{2304\alpha_1 \alpha_2}.
\end{equation}
Hence, Eq. \eqref{reduced-equation-1} yields
\begin{equation}
\eta^2 \simeq -\frac{1}{12\alpha_1} - \frac{1}{2}\left(-\frac{1}{9\alpha_1 \alpha_2}\right)^{1/4}.
\end{equation}
It turns out that the constraint $\eta \ll 1$ will require that 
\begin{equation}
\alpha_2 \sim -144\alpha_1^3 >0.
\end{equation}
 Additionally, the constraint $\zeta \gg 1$ will address the following constraint $|\alpha_1| \ll 1$ according to Eq. \eqref{reduced-equation-1} and therefore $\alpha_2 \ll 1$. This quantitive evaluation can be easily verified by the Fig. \ref{fig0}.
\subsection{A more general  case of Bianchi type I metric} \label{general case}
For now, ones might ask if a non-vanishing $\sigma_-$ leads to more general solutions of the SBR gravity. To address briefly this question, we will consider in this subsection a more general case with the following Bianchi type I metric given by \cite{barrow05,barrow06,Muller:2017nxg}
\begin{equation} \label{BI-metric-general}
ds^2 =-N^2(t)dt^2 +e^{2\alpha(t)-4\sigma_+(t)}dx^2 +e^{2\alpha(t)+2\sigma_+(t)+2\sqrt{3} \sigma_-(t)} dy^2 + e^{2\alpha(t)+2\sigma_+(t)-2\sqrt{3} \sigma_-(t)} dz^2.
\end{equation}
As a result, the corresponding Pontryagin density term ${P_4}$ still vanishes for this Bianchi type I metric, while the corresponding Ricci scalar and Gauss-Bonnet terms read
\begin{align}
 R  =&-\frac{6}{N^2} \left( \frac{\dot N}{N} \dot\alpha   -\ddot\alpha - 2\dot\alpha^2-\dot\sigma_+^2 -\dot\sigma_-^2\right),\\
{\cal G} =&-\frac{24}{N^5} \biggl\{ \left(\dot N - N \dot\alpha \right) \left(\dot \alpha -2\dot\sigma_+ \right) \left[\left(\dot\alpha+\dot\sigma_+\right)^2 -3\dot\sigma_-^2 \right] - N \left[\ddot\alpha \left(\dot\alpha^2-\dot\sigma_+^2 -\dot\sigma_-^2 \right) \right.\nonumber\\ 
&\left. -2\ddot\sigma_+ \left(\dot\alpha \dot\sigma_+ + \dot\sigma_+^2 -\dot\sigma_-^2 \right)  -2 \ddot\sigma_- \dot\sigma_- \left(\dot\alpha - 2\dot\sigma_+ \right) \right] \biggr\},
\end{align}
respectively. It is clear that $R$ and ${\cal G}$ will recover that defined in the previous subsection once the limits $\sigma_+ \to \sigma$ as well as $\sigma_ - \to 0$ are taken.
Now, using the similar ansatz,
  \begin{equation}
\alpha =\zeta t, ~\sigma_+= \eta_+ t,~\sigma_-= \eta_- t,
\end{equation}
we are able to obtain, after a lengthy calculation, the following set of algebraic equations from the corresponding Euler-Lagrange ones of $N$, $\alpha$, $\sigma_+$, and $\sigma_-$ as follows
\begin{align} \label{algebraic-equation-4}
& 18\alpha_1 \left(2\zeta^2 +\eta_+^2 +\eta_-^2 \right) \left(\eta_+^2 +\eta_-^2 \right)\nonumber\\
&+ 96\alpha_2 \zeta^2 \left(\zeta-2\eta_+ \right)^2 \left[ \left( \zeta  +\eta_+\right)^2 -3\eta_-^2 \right]^2   - \zeta^2 +\eta_+^2 +\eta_-^2 =0,\\
\label{algebraic-equation-5}
& 6\alpha_1 \left(2 \zeta^2+\eta_+^2 +\eta_-^2 \right)  \left(\eta_+^2 +\eta_-^2 \right) \nonumber\\
&-96 \alpha_2   \zeta^2 \left(\zeta-2\eta_+\right)^2 \left[ \left( \zeta  +\eta_+\right)^2 -3\eta_-^2 \right]^2 +\zeta^2 +\eta_+^2 +\eta_-^2 =0,\\
\label{algebraic-equation-6}
&\zeta \eta_+ \left[ 24\alpha_1 \zeta^2 +12\alpha_1 \left(\eta_+^2 +\eta_-^2 \right) +1 \right] =0,\\
\label{algebraic-equation-7}
&\zeta \eta_- \left[ 24\alpha_1 \zeta^2 +12\alpha_1 \left(\eta_+^2 +\eta_-^2 \right) +1 \right] =0.
\end{align} 
It is straightforward to see that these equations will reduce to Eqs. \eqref{algebraic-equation-1}, \eqref{algebraic-equation-2}, and \eqref{algebraic-equation-3}, respectively, once we set $\eta_+ \to \eta$ as well as $\eta_- \to 0$. As a result, both equations \eqref{algebraic-equation-6} and \eqref{algebraic-equation-7} implies a non-trivial relation,
\begin{align} \label{algebraic-equation-8}
2 \zeta^2 +\eta_+^2 +\eta_-^2  =-\frac{1}{12\alpha_1}.
\end{align}
This relation implies that it is impossible to have anisotropic solutions for a positive $\alpha_1$ even when $\sigma_-$ shows up in the Bianchi type I metric. However, a negative $\alpha_1$ can lead to the existence of anisotropic solutions.  Additionally, both Eqs. \eqref{algebraic-equation-4} and \eqref{algebraic-equation-5} will be reduced, thanks to this relation, to
\begin{equation} \label{algebraic-equation-9}
192\alpha_2 \zeta^2 \left(\zeta-2\eta_+ \right)^2 \left[ \left( \zeta  +\eta_+\right)^2 -3\eta_-^2 \right]^2 +\frac{1}{12\alpha_1}=0.
\end{equation}
It appears that we only have two independent equations \eqref{algebraic-equation-8} and \eqref{algebraic-equation-9} for three variables, $\zeta$, $\eta_+$, and $\eta_- $. Therefore, it seems to be difficult to derive exact analytical solutions of $\zeta$, $\eta_+$, and $\eta_-$, which depend only on the field parameters $\alpha_1$ and $\alpha_2$. However, we can figure out non-trivial relations among these three variables. For example, we can determine $\eta_+$ and $\eta_-$ as functions of $\zeta$ from these two equations. Indeed, we rewrite Eq.  \eqref{algebraic-equation-9} as follows
\begin{equation}  \label{algebraic-equation-10}
192\alpha_2 \zeta^2 \left(\zeta-2\eta_+ \right)^2 \left[\left( \zeta  +\eta_+\right)^2 +3\eta_+^2 +6\zeta^2 +\frac{1}{4\alpha_1} \right]^2 +\frac{1}{12\alpha_1}=0,
\end{equation}
with the help of Eq. \eqref{algebraic-equation-8}. This equation implies a non-trivial relation between $\eta_+$ and $\zeta$.  In principle, Eq. \eqref{algebraic-equation-10} can be solved to give its solutions of $\eta_+$, which depend on $\zeta$ as well as the field parameters $\alpha_1$ and $\alpha_2$. Then, plugging the obtained solutions of $\eta_+$ into Eq. \eqref{algebraic-equation-8} will yield the corresponding equation of $\eta_-$ and $\zeta$. Solving this equation will of course give us the desired solutions of $\eta_-$, which also depend on $\zeta$ as well as the field parameters $\alpha_1$ and $\alpha_2$. Once the value of $\zeta$ is estimated, the corresponding values of $\eta_+$ and $\eta_-$ will be determined accordingly. It appears that  the exact analytical forms of $\eta_+$ and $\eta_-$ in terms of $\zeta$, $\alpha_1$, and $\alpha_2$ would be complicated since Eq. \eqref{algebraic-equation-10} is highly nonlinear, i.e., a sixth-order equation of $\eta_+$. 

\begin{figure}[hbtp] 
\begin{center}
{\includegraphics[height=50mm]{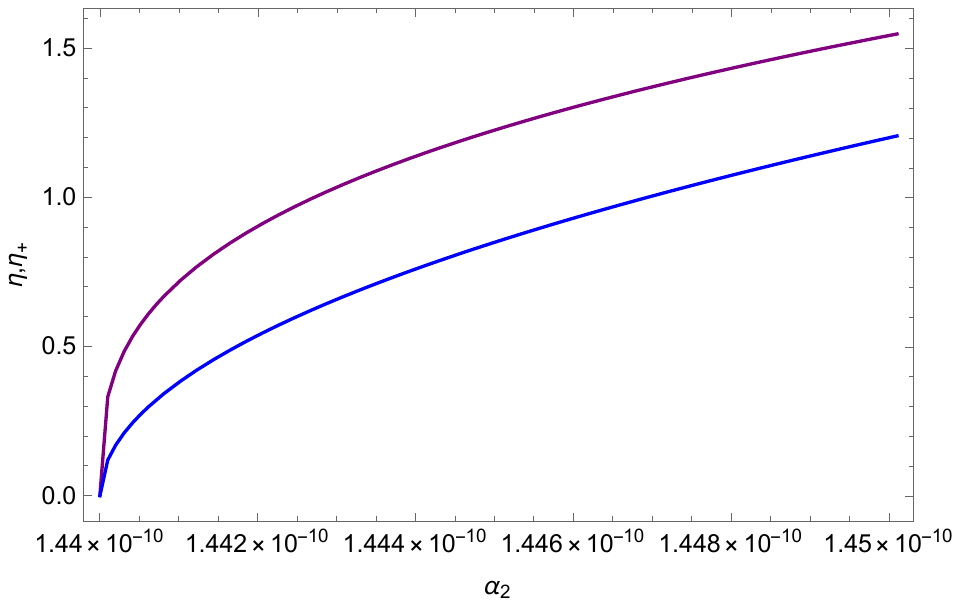}}\\
\caption{The upper purple curve is plotted for $\eta_+$, while the lower blue  curve is plotted for $\eta$. The field parameter $\alpha_1$ has been chosen as $\alpha_1= -10^{-4}$ for both curves.}
\label{fig2}
\end{center}
\end{figure}

However, we should note that we have concerned only with the inflationary phase, in which $\zeta \gg \eta_{\pm}$. Keeping this in mind, we are going to address approximated solutions of $\eta_+$ and $\eta_-$ during the inflationary phase.  First, 
we can have an approximation for $\zeta$, according to Eq.  \eqref{algebraic-equation-8}, as
\begin{equation}
\zeta^2  \simeq -\frac{1}{24\alpha_1}.
\end{equation}
Due to this result, Eq. \eqref{algebraic-equation-10} now becomes
\begin{equation}
192\alpha_2 \zeta^2 \left(\zeta-2\eta_+ \right)^2 \left[\left( \zeta  +\eta_+\right)^2 +3\eta_+^2 \right]^2 +\frac{1}{12\alpha_1}=0,
\end{equation}
which is clearly different from Eq. \eqref{reduced-equation-2}. This means that $\eta_+\neq \eta$ as expected. Furthermore, it can be reduced to
\begin{equation}
96\alpha_2 \left(\zeta^3 -8\eta_+^3\right)^2  -1 =0,
\end{equation}
due to the relation $\alpha_1 =-1/(24\zeta^2)$. As a result, solving this equation gives
\begin{equation}
\left(\eta_+^3\right)^\pm = \frac{\zeta^3}{8} \pm \frac{1}{32 \sqrt{6\alpha_2}}.
\end{equation}
It appears that only 
\begin{equation}
\eta_+ \simeq \eta_+^- = \left(\frac{\zeta^3}{8} - \frac{1}{32 \sqrt{6\alpha_2}}\right)^{1/3}
\end{equation} is our desired approximated solution. Similar to the  solution found in the previous subsection, $\eta_+ \ll 1$ if  $\alpha_2 \sim -144\alpha_1^3$. 
Given this solution, $\eta_-$ can be defined, according to Eq. \eqref{algebraic-equation-8}, to be
\begin{equation}
\eta_-^2 \simeq -\frac{1}{12\alpha_1} -2\zeta^2 -\left(\frac{\zeta^3}{8} - \frac{1}{32 \sqrt{6\alpha_2}}\right)^{2/3}.
\end{equation} 
To see the effect of non-vanishing $\eta_-$ on the value of $\eta_+$, we will compare the value of $\eta$ found in the previous subsection with the value of $\eta_+$ found in this subsection. According to Fig. \ref{fig2}, we observe that when $\alpha_2 \to -144\alpha_1^3$ both $\eta_+$ and $\eta$ tend to be much smaller than one as expected. More interestingly, the value of $\eta_+$ is always slightly larger than $\eta$. This result together with Eqs. \eqref{reduced-equation-1} and \eqref{algebraic-equation-8}  indicate that the value of $\zeta$ derived in the case of non-vanishing $\eta_-$ must be slightly smaller than that derived in the case of vanishing $\eta_-$.  Due to these analysis, it is safe to conclude that the effect of non-vanishing $\sigma_-$ is not significant during the inflationary phase. Therefore, we can turn off $\sigma_-$ for simplicity.  It should be noted again that the choice of vanishing $\sigma_-$ can be found in a number of papers on anisotropic inflation, e.g., Refs.  \cite{Kanno:2010nr,Do:2011zza}, where exact anisotropic solutions have been figured out. 
\section{Stability analysis: Dynamical system method} \label{sec3}
In this section, we would like to address an important question that whether the obtained solutions are stable or not. To do this, we will transform the field equations into the corresponding dynamical system, similar to the previous works on fourth-order gravity models \cite{Do:2020vdc,barrow05,barrow06,Muller:2017nxg}, by introducing dimensionless dynamical variables such as
 \begin{align}
& B=\frac{1}{\dot\alpha^2},\nonumber\\
&Q=\frac{\ddot\alpha}{\dot\alpha^2},\nonumber\\
& Q_2 =\frac{\alpha^{(3)}}{\dot\alpha^3}, \nonumber\\
& \Sigma =\frac{\dot\sigma}{\dot\alpha}, \nonumber\\
&\Sigma_1 =\frac{\ddot\sigma}{\dot\alpha^2}, \nonumber\\
&\Sigma_2 =\frac{\sigma^{(3)}}{\dot\alpha^3}, 
 \end{align}
 here the Hubble constant is given by $H=\dot\alpha$. Note that the notations of dynamical variables come from Refs. \cite{barrow05,barrow06}.  As a result, the corresponding set of autonomous equations of dynamical variables are given by 
 \begin{align}
  \label{Dyn-1}
 B' &= -2QB,\\
  \label{Dyn-3}
 Q' &=Q_2 -2Q^2,\\
 \label{Dyn-6}
 Q_2 '&= \frac{\alpha^{(4)}}{\dot\alpha^4}-3Q Q_2,\\
  \label{Dyn-4}
 \Sigma' &= \Sigma_1 - Q\Sigma ,\\
  \label{Dyn-5}
 \Sigma_1'&= \Sigma_2 -2Q \Sigma_1,\\
  \label{Dyn-7}
 \Sigma_2'&= \frac{\sigma^{(4)}}{\dot\alpha^4}-3Q \Sigma_2,
 \end{align}
 where $' \equiv d/d\tau$ with $\tau =\int \dot\alpha dt$ being the dynamical time variable. It should be noted that the terms ${\alpha^{(4)}}/{\dot\alpha^4}$ and ${\sigma^{(4)}}/{\dot\alpha^4}$ in Eqs. \eqref{Dyn-6} and \eqref{Dyn-7} can be determined from the field equations \eqref{alpha-field-eq-BI} and \eqref{sigma-field-eq-BI}, which can be reformed in terms of the dynamical variables such as
 \begin{align} \label{Dyn-8}
 &6\alpha_1 B^2 \left[ 2\frac{\alpha^{(4)}}{\dot\alpha^4}+12 Q_2 +4\Sigma \Sigma_2 +9Q^2 +4\Sigma_1^2 +2Q\left(2\Sigma^2+9 \right) +8\Sigma \Sigma_1 +3\Sigma^2\left(\Sigma^2 +2\right) \right] \nonumber\\
& + 96\alpha_2 \biggl\{ 2 \frac{\alpha^{(4)}}{\dot\alpha^4} \left( 1 -\Sigma^2 \right)^2 -4\frac{\sigma^{(4)}}{\dot\alpha^4} \Sigma \left(1-\Sigma \right) \left(1+\Sigma \right)^2  \nonumber\\
&  +4Q_2 \left(1-\Sigma^2 \right) \left[4 Q -4\Sigma \Sigma_1 +3 \left(1-\Sigma^2 \right) \right]  \nonumber\\
 &  -4 \Sigma_2 \left(1+\Sigma \right) \left[Q \Sigma \left(5-3\Sigma \right) + \Sigma_1 \left(3+3\Sigma -8\Sigma^2 \right) +  \Sigma \left(1+\Sigma \right) \left(5-5\Sigma+2\Sigma^2  \right) \right] \nonumber\\
 &+4 Q^3  \left(3-\Sigma^2 \right) -8\Sigma_1^3 \left(1-\Sigma -3\Sigma^2 \right)    -Q^2 \left(24\Sigma \Sigma_1 -45-9\Sigma^4 +54 \Sigma^2 \right) \nonumber\\
 & -4 \Sigma_1^2 \left[Q \left(5+4\Sigma-7\Sigma^2 \right) +5+7\Sigma^4 +2\Sigma \left(1-\Sigma\right) \left(5+2\Sigma \right) \right] \nonumber\\
 & -4 Q\Sigma_1\Sigma  \left(1+\Sigma\right)\left(23 -14 \Sigma -\Sigma^2 \right) \nonumber\\
 &   +2Q \left(1+\Sigma\right)^2 \left(5-10\Sigma +15 \Sigma^2 -10 \Sigma^3 -4\Sigma^4 \right)  \nonumber\\
&  - 12 \Sigma \Sigma_1 \left(1+\Sigma\right)^2 \left(1-\Sigma +4\Sigma^2 \right)  -3 \left(1+\Sigma\right)^4 \left(1-2\Sigma \right)^2 \biggr\} +B^3 \left( 2Q +3 +3\Sigma^2\right) =0,
\end{align}
\begin{align}
 \label{Dyn-9}
 &12 \alpha_1 B^2 \left[ Q_2 \Sigma +Q \left(\Sigma_1 +7\Sigma \right) +\Sigma_1 \left(2+3\Sigma^2 \right) +3\Sigma \left( 2+\Sigma^2 \right) \right] \nonumber\\
 & +192 \alpha_2 \biggl\{ \frac{\alpha^{(4)}}{\dot\alpha^4} \Sigma \left( 1-\Sigma \right) \left(1+\Sigma \right)^2 -2\frac{\sigma^{(4)}}{\dot\alpha^4}\Sigma^2 \left(1+\Sigma \right)^2  \nonumber\\
 & +Q_2 \left(1+\Sigma \right)\left[Q\Sigma \left(7-\Sigma \right) +\Sigma_1 \left(1+\Sigma -8\Sigma^2 \right) +\Sigma \left(1+\Sigma \right) \left(7-7\Sigma -2\Sigma^2 \right) \right]  \nonumber\\
 & -4\Sigma_2 \Sigma \left(1+\Sigma \right) \left[2Q\Sigma +2\Sigma_1 \left(1+2\Sigma \right) +3\Sigma \left(1+\Sigma \right) \right] \nonumber\\
 &  +2 Q^3 \Sigma \left(2+\Sigma\right) -2\Sigma_1^3 \left(1+6\Sigma +6\Sigma^2 \right) \nonumber\\
 & +2 Q^2 \left[\Sigma_1 \left(1+2\Sigma -2\Sigma^2\right)+\Sigma \left(1+\Sigma\right) \left(11-2\Sigma -4\Sigma^2 \right) \right]  \nonumber\\
 & -6 \Sigma_1^2 \Sigma \left[Q \left(2+3\Sigma\right) +3 \left(1+\Sigma \right) \left(1+2\Sigma \right) \right] \nonumber\\
 & + Q \left(1+\Sigma\right) \left[4\Sigma_1 \left(1+4\Sigma \right)\left(1-3\Sigma -\Sigma^2 \right) +6\Sigma \left(1+\Sigma\right) \left(2-2 \Sigma - \Sigma^2 \right) \right] \nonumber\\
 & -18 \Sigma_1 \Sigma^2 \left(1+\Sigma \right)^2 \biggr\}   + B^3 \left(\Sigma_1 +3\Sigma \right)=0.
 \end{align}
In other words, solving Eqs. \eqref{Dyn-8} and \eqref{Dyn-9} will yield the corresponding expression of ${\alpha^{(4)}}/{\dot\alpha^4}$ and ${\sigma^{(4)}}/{\dot\alpha^4}$, which will be purely defined in terms of the dynamical variables introduced above. Consequently, we will be able to obtain the corresponding complete dynamical system of autonomous equations \eqref{Dyn-1}, \eqref{Dyn-3}, \eqref{Dyn-6}, \eqref{Dyn-4}, \eqref{Dyn-5}, and \eqref{Dyn-7}, which are the first-order differential equations of dynamical variables $B$, $Q$, $Q_2$, $\Sigma$, $\Sigma_1$, and $\Sigma_2$. The reason we have skipped showing the detailed expressions of ${\alpha^{(4)}}/{\dot\alpha^4}$ and ${\sigma^{(4)}}/{\dot\alpha^4}$ is due to their lengthy display, which are the consequence of the high nonlinearity of Eqs. \eqref{Dyn-8} and \eqref{Dyn-9}. However, we will show later that Eqs. \eqref{Dyn-8} and \eqref{Dyn-9} will turn out to be easily solved for fixed points. We should note that it is unnecessary to introduce three variables, $\Sigma$, $\Sigma_1$, and $\Sigma_2$ for the isotropic FLRW metric having $\sigma(t) =0$. However, we would like to define here the general dynamical system of six dynamical variables, $B$, $Q$, $Q_2$, $\Sigma$, $\Sigma_1$, and $\Sigma_2$, since we have planned to seek both isotropic and anisotropic fixed points.
 
 Besides, the Friedmann equation \eqref{N-field-eq-BI} can also be rewritten in terms of the dynamical variables to be
 \begin{align} \label{Dyn-10}
 &6\alpha_1 B^2 \left[2 Q_2 - Q \left(Q-6+4\Sigma^2 \right) + 4 \Sigma \Sigma_1 -3\Sigma^2 \left(2+\Sigma^2 \right)\right] \nonumber\\
& +96 \alpha_2 \left(1+\Sigma \right)^2 \left[ 2Q_2 \left(1-2\Sigma \right)\left(1-\Sigma^2\right) -4\Sigma_2\Sigma \left(1+\Sigma\right) \left(1-2\Sigma\right) +Q^2 \left(3-6\Sigma -\Sigma^2 \right) \right. \nonumber\\
 &\left. -4\Sigma_1^2  \left(1-3\Sigma^2 \right) -4 Q \Sigma_1 \Sigma \left(1-3\Sigma \right) + 2Q \left(1+\Sigma \right) \left(1-2\Sigma \right)\left(3-3\Sigma -2\Sigma^2 \right) \right.\nonumber\\
&\left. -8\Sigma_1 \Sigma \left(1+\Sigma \right) \left(1-2\Sigma \right) -\left(1+\Sigma\right)^2 \left(1-2\Sigma \right)^2 \right] +B^3 \left(1-\Sigma^2\right) =0,
 \end{align}
 which the found fixed points must satisfy accordingly. In other words, Eq. \eqref{Dyn-10} plays as the unavoidable constraint equation of the dynamical system of autonomous equations \eqref{Dyn-1}, \eqref{Dyn-3}, \eqref{Dyn-6}, \eqref{Dyn-4}, \eqref{Dyn-5}, and \eqref{Dyn-7}.
 \subsection{Isotropic fixed point}
 Given the above dynamical system, we are going to seek its isotropic fixed points. Mathematically, fixed points are solutions of the following set of equations, 
 \begin{equation}
B'=Q'=Q_2'=\Sigma' =\Sigma_1'=\Sigma_2' =0.
\end{equation}
 Additionally, isotropic fixed points will correspond to
 \begin{equation}
 \Sigma =\Sigma_1=\Sigma_2=0.
 \end{equation}
 Consequently, we have
 \begin{equation}
 Q=Q_2=\frac{\alpha^{(4)}}{\dot\alpha^4}=\frac{\sigma^{(4)}}{\dot\alpha^4}=0.
 \end{equation}
 As a result, Eq. \eqref{Dyn-9} is automatically satisfied, while Eqs. \eqref{Dyn-8} and \eqref{Dyn-10} both reduce to
 \begin{equation}
 B^3 -96 \alpha_2 =0,
 \end{equation}
 which is consistent with  Eq. \eqref{equation-of-zeta-iso}. Integrating this equation leads to an exact solution,
 \begin{equation}
 \alpha(t)= \zeta t,
 \end{equation}
where $\zeta$ has been defined in Eq. \eqref{sol-of-zeta-FLRW}. This result  emphasizes that this isotropic fixed point is equivalent to the de Sitter solution found in the previous section.
 \subsection{Anisotropic fixed point}
 We now look for anisotropic fixed points with $\Sigma \neq 0$. It turns out that $B'=0$ implies that $Q=0$. Consequently, the other equations, $Q'=0$, $\Sigma'=0$, and $\Sigma_1'=0$, all admit $Q_2=\Sigma_1=\Sigma_2=0$. Furthermore, the remaining equations, $Q_2'=0$ and $\Sigma_2' =0$, can be solved to give
 \begin{equation}
 \frac{\alpha^{(4)}}{\dot\alpha^4}=\frac{\sigma^{(4)}}{\dot\alpha^4}=0.
 \end{equation}
 Thanks to these solutions, Eqs. \eqref{Dyn-8}  and \eqref{Dyn-9} reduce to
 \begin{align} \label{anisotropic-fixed-point-1}
 6\alpha_1 B^2 \Sigma^2 \left(\Sigma^2 +2 \right)- 96\alpha_2  \left(2\Sigma-1 \right)^2  \left(\Sigma+1\right)^4 +B^3 \left(\Sigma^2+1 \right) =0,\\
 \label{anisotropic-fixed-point-2}
 12 \alpha_1   \left(\Sigma^2+2 \right) + B =0.
  \end{align} 
  Moreover, the constraint equation \eqref{Dyn-10} also becomes as
  \begin{equation}
  \label{anisotropic-fixed-point-3}
  18 \alpha_1 B^2 \Sigma^2 \left(\Sigma^2+2 \right) +96\alpha_2 \left(2\Sigma-1 \right)^2 \left(\Sigma +1 \right)^4 +B^3 \left(\Sigma^2-1 \right) =0.
  \end{equation}
  Interestingly, Eqs. \eqref{anisotropic-fixed-point-1} and \eqref{anisotropic-fixed-point-3} both can be further simplified to the same equation,
  \begin{equation} \label{anisotropic-fixed-point-4}
  192 \alpha_2  \left(2\Sigma-1 \right)^2  \left(\Sigma+1\right)^4 +\frac{B^4}{12\alpha_1} =0,
  \end{equation}
  with the help of Eq. \eqref{anisotropic-fixed-point-2}. It appears that Eqs. \eqref{anisotropic-fixed-point-2} and \eqref{anisotropic-fixed-point-4} will reduce to  Eqs. \eqref{reduced-equation-1} and \eqref{reduced-equation-2}, respectively, once we set $\alpha(t) =\zeta t$ and $\sigma(t) =\eta t$. Therefore, we can conclude that the anisotropic fixed point is equivalent to the exact Bianchi type I solution found in the previous section.
 \subsection{Stability analysis of isotropic fixed point}
Up to now, we have figured out equations for the existence of both isotropic and anisotropic fixed points to the dynamical system of the SBR gravity. These equations are indeed consistent with that derived in the previous section for the exact exponential solutions. In other words, the found fixed points are indeed equivalent to the exponential solutions of field equations. Therefore, the stability of fixed points will tell us the stability of the exponential solutions. It is important to note again that the isotropic fixed point, or equivalently the de Sitter solution, has been shown to be unstable against field perturbations by a different approach in Ref. \cite{Ketov:2022zhp}.

In order to investigate whether the isotropic fixed point is stable or not, we will perturb the autonomous equations around this fixed point as follows
\begin{align}
\delta B' &=-2B\delta Q,\\
\delta Q' &= \delta Q_2,\\
\delta Q_2' &= \delta \left(\frac{\alpha^{(4)}}{\dot\alpha^4} \right),
\end{align}
here $\delta \left(\frac{\alpha^{(4)}}{\dot\alpha^4} \right)$ is defined from the perturbed equation,
\begin{align}
&12 \left(\alpha_1 B^2 + 16\alpha_2 \right) \delta \left(\frac{\alpha^{(4)}}{\dot\alpha^4} \right)  +2 \left(54 \alpha_1 B^2 + 480\alpha_2 +B^3 \right) \delta Q  \nonumber\\
&+ 6 \left( 12 \alpha_1 B^2 + 192 {\alpha_2}  \right)\delta Q_2 +9B^2 \delta B =0,
\end{align}
to be
\begin{equation}
\delta \left(\frac{\alpha^{(4)}}{\dot\alpha^4} \right) = -\frac{1}{6\left(\alpha_1 B^2 + 16\alpha_2 \right)} \left[ \left(B^3-384\alpha_2   \right)\delta Q + 18\left(\alpha_1 B^2- 16\alpha_2   \right)\delta Q_2 \right],
\end{equation}
with the help of the perturbed Friedmann equation,
\begin{equation}
4\left(\alpha_1 B^2 + 16 \alpha_2 \right) \left(3\delta Q+ \delta Q_2  \right) +B^2 \delta B=0. 
\end{equation}
Again, it should be noted that Eqs. \eqref{Dyn-4}, \eqref{Dyn-5}, and \eqref{Dyn-7} of three dynamical variables $\Sigma$, $\Sigma_1$, and $\Sigma_2$, have been ignored for now since they are unimportant to the stability of the isotropic fixed point. 
By taking exponential perturbations, 
\begin{equation}
\delta B, ~\delta Q, ~\delta Q_2 \sim \exp[\mu\tau],
\end{equation}
we are able to obtain the following quadratic equation of $\mu$ from the perturbed equations,
\begin{equation} \label{quadratic-equation}
 \left(\alpha_1 B^2+16 \alpha_2  \right) \mu^2 +3 \left(\alpha_1 B^2 -16 \alpha_2  \right) \mu -48\alpha_2 =0,
\end{equation}
here the isotropic fixed point solution, $B^3 =96\alpha_2$, has been used, and the trivial solution, $\mu_1 =0$, has been ignored. It is clear that this quadratic equation of $\mu$ will always admit at least one positive root, $\mu >0$, if the coefficient of second degree term is always positive definite, i.e., 
\begin{equation}
\alpha_1 B^2+16 \alpha_2  >0,
\end{equation}
or equivalently,
\begin{equation}
\alpha_1 > -\left(\frac{4\alpha_2}{9}\right)^{1/3},
\end{equation}
 due to the fact that the coefficient of zeroth degree term is always negative definite, i.e., $-48\alpha_2<0$. It is straightforward to see that the positivity of $\alpha_1$, which has been required in the pure Starobinsky model, will lead to the instability of the  isotropic fixed point. This means that the corresponding de Sitter solution found in the previous section is indeed unstable for the positive $\alpha_1$. This conclusion is really consistent with the investigation presented in Ref. \cite{Ketov:2022zhp}. However, the result in the previous section indicates that $\alpha_1$ does not contribute to the value of the de Sitter solution. This means that $\alpha_1$ may be assumed to be a free parameter. Hence, a modified SBR model with a negative $\alpha_1$ will also admit a de Sitter solution. Note that the negativity of $\alpha_1$ can be done by the flipping $\alpha_1 \to \alpha_1 = -(6m^2)^{-1}<0$.  Very interestingly, if $\alpha_1$ could be negative enough such that
 \begin{equation}
\alpha_1 < -\left(\frac{4\alpha_2}{9}\right)^{1/3},
 \end{equation}
 then all coefficients of the above quadratic equation of $\mu$ will turn out to be negative. This ensures that this equation will no longer admit any positive roots $\mu >0$. And therefore, the corresponding de Sitter solution will be stable accordingly. However, the negativity of $\alpha_1$ would definitely breaks a smooth connection between the modified SBR model with the pure Starobinsky model. In other words, the modified SBR gravity model with a negative $\alpha_1$  will not recover the pure Starobinsky model with a positive $\alpha_1$ once the limit $\alpha_2 \to 0$ is taken. Furthermore, the negativity of $\alpha_1$ would modify significantly the CMB predictions of the modified SBR model compared to that of the pure Starobinsky model.  It is worth noting  that a stable de Sitter inflationary solution, which may be a basis of eternal inflation \cite{Guth:2007ng}, seems to be unsuitable for describing the inflationary phase of early universe due to the so-called graceful exit problem \cite{Elizalde:2014xva,Pozdeeva:2019agu}. However, if we could figure out a suitable mechanism for the graceful exit, the stable de Sitter inflationary solution could be our desirable one \cite{Elizalde:2014xva}. Unfortunately, figuring out such mechanism seems to be a difficult task.  Cosmologically, the graceful exit is a necessary transition happening at the end of  inflationary phase in order to ensure a smooth connection between an early inflationary phase and a late time expanding FLRW phase. See Ref. \cite{Brustein:1994kw} for an interesting paper on this problem.  Therefore, gravity models admitting no de Sitter solution or unstable de Sitter solutions during the inflationary phase seem to have a better chance of being realistic inflationary models. It turns out that quasi-de Sitter solutions may happen in these models. On the other hand, gravity models admitting stable de Sitter solutions seem to be suitable for the late time acceleration of our universe \cite{Pozdeeva:2019agu}. Due to the existence of unstable de Sitter solution as shown in Ref. \cite{Ketov:2022zhp} and confirmed in this subsection, the original SBR gravity model could be relevant to the inflationary phase. Indeed, CMB predictions such as the tilt $n_s$ of scalar (curvature) perturbations and the tensor-to-scalar ratio $r$ of the SBR gravity model have been worked out in a recent paper \cite{Ketov:2022zhp}. As a result, there are small gaps between predicted values of the pure Starobinsky gravity model and the SBR gravity model, provided that $\beta$ is very small ($0\leq \beta \leq 3.9 \times 10^{-6}$). Additionally, the quantum correction to the Starobinsky inflation model tends to decrease both $n_s$ and $r$ because of the positive $\beta$. It appears that cosmological implications of  the modified SBR gravity model having a negative $\alpha_1$ require an independent systematic investigation. We will therefore leave this issue to our future study, whose results will be presented elsewhere.
 \subsection{Stability analysis of anisotropic fixed point}
 Next, we would like to investigate the stability of the obtained anisotropic fixed point with $\Sigma \neq 0$. As a result, the following perturbed equations are given by
 \begin{align}
\delta B' &=-2B\delta Q,\\
\delta Q' &= \delta Q_2,\\
\delta Q_2' &= \delta \left(\frac{\alpha^{(4)}}{\dot\alpha^4} \right),\\
\delta \Sigma' &= \delta\Sigma_1 -\Sigma \delta Q ,\\
\delta \Sigma_1' &= \delta \Sigma_2 ,\\
\delta \Sigma_2' &= \delta\left(\frac{\sigma^{(4)}}{\dot\alpha^4} \right),
\end{align}
here $\delta\left(\frac{\alpha^{(4)}}{\dot\alpha^4} \right) $ and $\delta \left(\frac{\sigma^{(4)}}{\dot\alpha^4}\right) $ will be figured out from two perturbed equations,
\begin{align} \label{X}
&6\alpha_1 B^2 \left[ 2 \delta \left(\frac{\alpha^{(4)}}{\dot\alpha^4}\right) +12\delta Q_2 +4\Sigma \delta \Sigma_2  +2\left(2\Sigma^2+9\right)\delta Q +8\Sigma \delta\Sigma_1 +12\Sigma \left(\Sigma^2 +1 \right)\delta \Sigma    \right]\nonumber\\
&+36\alpha_1 B \Sigma^2\left(\Sigma^2+2 \right)\delta B \nonumber\\
&+ 96\alpha_2 \biggl\{ 2 \left(\Sigma^2- 1 \right)^2 \delta \left(\frac{\alpha^{(4)}}{\dot\alpha^4}\right) + 4  \Sigma \left(\Sigma-1 \right) \left(\Sigma+1 \right)^2 \delta \left(\frac{\sigma^{(4)}}{\dot\alpha^4} \right) \nonumber\\
&+12 \left(\Sigma^2-1 \right)^2 \delta Q_2  - 4 \Sigma \left(\Sigma+1 \right)^2 \left(2\Sigma^2-5\Sigma +5\right) \delta\Sigma_2 \nonumber\\
& -2 \left(\Sigma+1\right)^2 \left(4\Sigma^4+10 \Sigma^3-15 \Sigma^2   +10\Sigma-5 \right) \delta Q \nonumber\\
& -12\Sigma \left(\Sigma+1 \right)^2 \left(4\Sigma^2 -\Sigma+1 \right) \delta \Sigma_1  -36\Sigma \left(2\Sigma-1 \right) \left(\Sigma+1 \right)^3  \delta \Sigma \biggr\} \nonumber\\
&+9B^2 (\Sigma^2+1)\delta B +2B^3\left(\delta Q + 3\Sigma \delta \Sigma \right)=0,
\end{align}
\begin{align}
\label{Y}
&12\alpha_1 B^2 \left[\Sigma \delta Q_2 +7\Sigma \delta Q +2\delta \Sigma_1 +3\Sigma^2 \delta \Sigma_1 +3\left(3\Sigma^2+2 \right)\delta \Sigma \right] \nonumber\\
&+72\alpha_1 B \Sigma \left(\Sigma^2+2 \right) \delta B \nonumber\\
& - 192\alpha_2 \biggl\{ \Sigma \left(\Sigma-1 \right) \left(\Sigma+1 \right)^2  \delta \left(\frac{\alpha^{(4)}}{\dot\alpha^4}\right) + 2\Sigma^2 \left(\Sigma+1 \right)^2 \delta \left(\frac{\sigma^{(4)}}{\dot\alpha^4} \right)  \nonumber\\
& + \Sigma \left(\Sigma+1 \right)^2 \left(2\Sigma^2 +7\Sigma -7 \right)\delta Q_2  + 12 \Sigma^2 \left(\Sigma+1 \right)^2 \delta \Sigma_2  \nonumber\\
& + 6\Sigma \left(\Sigma+1\right)^2 \left(\Sigma^2+2\Sigma-2 \right) \delta Q 
 + 18\Sigma^2 \left(\Sigma+1 \right)^2 \delta\Sigma_1  \biggr\} \nonumber\\
 & + 9B^2 \Sigma \delta B +B^3 \left(\delta\Sigma_1 +3\delta\Sigma \right) =0.
\end{align}
In order to simplify the above perturbed equations, we will need the perturbed Friedmann equation,
\begin{align}
&6\alpha_1 B^2 \left[ 2\delta Q_2 - 2\left(2\Sigma^2 -3 \right) \delta Q + 4\Sigma \delta\Sigma_1 -12\Sigma \left(\Sigma^2 +1 \right)\delta\Sigma \right]  -36 \alpha_1 B\Sigma^2 \left(\Sigma^2+2\right)\delta B \nonumber\\
&+96 \alpha_2 \left(\Sigma+1 \right)^2  \Bigl\{ 2 \left(2\Sigma-1 \right)\left(\Sigma^2-1 \right) \delta Q_2 + 4 \Sigma \left(\Sigma+1 \right) \left(2\Sigma-1\right)\delta\Sigma_2   \nonumber\\
& +2\left(\Sigma+1 \right)\left(2\Sigma-1\right)\left(2\Sigma^2+3\Sigma-3 \right)\delta Q  +8\Sigma \left(\Sigma+1 \right)\left(2\Sigma -1 \right) \delta\Sigma_1 \nonumber\\
& -12\Sigma \left(\Sigma+1 \right) \left(2\Sigma-1 \right) \delta \Sigma \Bigr\} - 3B^2 \left(\Sigma^2-1 \right)\delta B -2B^3 \Sigma \delta \Sigma =0.
\end{align}
Before going to seek eigenvalues, we should note that 
\begin{equation} \label{constraint-of-Sigma}
|\Sigma| = \left |\frac{\eta}{\zeta} \right| \ll 1,
\end{equation}
 is a reasonable constraint for ensuring that the obtained anisotropic inflationary solution would be consistent with observational data due to a small anisotropy. It should be noted that $\eta$ can be either positive or negative definite but its absolute value should be much smaller than $\zeta$, which mainly governs the expansion of the inflationary phase, because we would like to have the corresponding inequalities,
 \begin{equation}
 \zeta -2 \eta \simeq \zeta \gg 1,~ \zeta+\eta \simeq \zeta \gg 1.
 \end{equation}
It is worth noting that some CMB anomalies such as the hemispherical asymmetry and the cold spot, which have been confirmed by the Planck \cite{Schwarz:2015cma}, have provided a hint that an anisotropic inflationary phase having a stable small spatial anisotropy might have a chance to happen in the early time. In particular, if the obtained anisotropic inflation was stable against perturbations then its small anisotropy would remain before and after a graceful exit. And this remaining small anisotropy, which is indeed beyond the prediction of the cosmic no-hair conjecture, may be relevant to the existence of the CMB anomalies.
 
On the other hand, ones might think of a more complicated scenario, in which $|\eta|$ is not much smaller than $\zeta$, i.e., the constraint shown in Eq. \eqref{constraint-of-Sigma} would not be fulfilled. For this case, we would have an anisotropic inflation with a large spatial anisotropy. In order to be compatible with the current  observational data of the Planck, however, this anisotropy should be unstable, i.e., decays with time. However, it is quite complicated to see when this anisotropy becomes small enough to be consistent with the observation. In this paper, therefore, we will focus only on a simpler scenario, in which the constraint  \eqref{constraint-of-Sigma} is assumed to be valid.
  
  It turns out that this constraint helps us to reduce the lengthy calculations.  As a result, $\delta\Sigma$ can be solved from this equation to be
\begin{equation}
\delta\Sigma \simeq \frac{1}{104\alpha_1 \Sigma} \left( \delta B -36\alpha_1 \delta Q -12\alpha_1 \delta Q_2 +72 \alpha_1 \Sigma \delta \Sigma_1 +32\alpha_1 \Sigma \delta \Sigma_2  \right) ,
\end{equation}
here we have used the approximations that $B\simeq -24\alpha_1$ as well as $\alpha_2 \simeq -144 \alpha_1^3$ for the anisotropic inflationary solution. Plugging this $\delta \Sigma$ into Eqs. \eqref{X} and \eqref{Y} will lead to the corresponding expressions,
\begin{align}
\delta\left(\frac{\alpha^{(4)}}{\dot\alpha^4} \right)  & \simeq \frac{1}{4\alpha_1}\left( \delta B-212\alpha_1 \delta Q -72\alpha_1 \delta Q_2 +392 \alpha_1 \Sigma \delta \Sigma_1 +120  \alpha_1 \Sigma \delta\Sigma_2 \right),\\
\delta\left(\frac{\sigma^{(4)}}{\dot\alpha^4} \right)  & \simeq  \frac{1}{32\alpha_1 \Sigma }\left( 3\delta B -684 \alpha_1 \delta Q -180\alpha_1 \delta Q_2 +1280 \alpha_1 \Sigma \delta \Sigma_1 +288 \alpha_1 \Sigma \delta \Sigma_2 \right).
\end{align}
Now, we take the exponential perturbations, 
\begin{equation}
\delta B, ~\delta Q, ~\delta Q_2, ~\delta \Sigma,~\delta \Sigma_1,~\delta\Sigma_2 \sim \exp[\mu\tau].
\end{equation}
As a result, the corresponding quintic equation of $\mu$ is given by
\begin{equation}
\mu  \left(\mu+3 \right) \left(\mu+4 \right) \left(\mu-1\right) \left(4\mu^2 +12 \mu -13 \right) =0,
\end{equation}
Beside the trivial root $\mu_1=0$ and three negative roots $\mu_2=-3$, $\mu_3=-4$, and $\mu_4=- \left(3+\sqrt{22} \right)/2$, it is clear that this equation always admits two positive roots, one is $\mu_5 =1$, and the other $\mu_6 =-\left(3-\sqrt{22} \right)/2$. This means that the anisotropic fixed point is always unstable against perturbations.  It should be noted that the instabilities of both isotropic and anisotropic solutions found in this section are purely at the classical level and therefore have nothing to do with the Ostrogradsky ghost. For other classical instabilities of the Bianchi type I inflationary solution of higher-than-two order gravity models, e.g., the Einsteinian cubic gravity \cite{Bueno:2016xff}, one can see Ref. \cite{Pookkillath:2020iqq}.
\section{Conclusions}\label{final}
We have studied the so-called Starobinsky-Bel-Robinson gravity model, which is a superstring-inspired quantum correction to the Starobinsky model of inflation \cite{Starobinsky:1980te} constructed by Ketov recently \cite{Ketov:2022lhx}. In particular, we have derived exactly isotropic and/or anisotropic exponential solutions to this SBR model as well as its modified version. For the de Sitter solution, we have seen that only $\alpha_2$ contributes to the value of scale factor. However, for the Bianchi type I solutions, both $\alpha_1$ and $\alpha_2$ govern the value of scale factors. More interestingly, a positive $\alpha_1$ will not lead to the existence of spatial anisotropies, while a negative $\alpha_1$ will cause the appearance of anisotropic solution. In order to examine the stability of the obtained solutions during the inflationary phase, we have transformed the field equations into the corresponding dynamical system. As a result, (an)isotropic fixed points of this system are equivalent to the exponential ones of field equations. Interestingly, we are able to confirm the result, which was first obtained in the previous paper \cite{Ketov:2022zhp}, that the de Sitter inflation is indeed unstable, by showing that the corresponding  isotropic fixed point is unstable against perturbations. Moving on to the anisotropic fixed point, we have also obtained the similar conclusion that the corresponding anisotropic inflation is unstable against perturbations. All these results support an observation that the Starobinsky model is very sensitive with the inclusion of other higher-order corrections. Interestingly, we have obtained a very interesting point that if we assumed $\alpha_1$ as a negative parameter such that $\alpha_1  < - \left(4\alpha_2 /9 \right)^{1/3}$, which would of course conflict with the Starobinsky model, then a stable de Sitter inflationary solution would exist. However, the modified SBR model with a negative $\alpha_1$ would not reduce to the Starobinsky model in the limit $\alpha_2 \to 0$. Additionally, the corresponding CMB predictions of the modified SBR model with a negative $\alpha_1$ would be different from that of the Starobinsky model with a positive $\alpha_1$. More importantly, this stable de Sitter inflationary solution, which may be a basis of eternal inflation \cite{Guth:2007ng}, would need a suitable mechanism for its graceful exit  in order to be more realistic \cite{Elizalde:2014xva}. Unfortunately, figuring out such mechanism seems to be a difficult task. In some other gravity models, a stable de Sitter solution has been preferred to use for the late time cosmic acceleration rather than the early time cosmic inflation \cite{Pozdeeva:2019agu}. The instability of de Sitter inflationary solution of the original SBR model indicates that this model could be relevant to the inflationary phase \cite{Ketov:2022zhp}.  A detailed investigation for cosmological implications of the stable de Sitter solution within the modified version of the SBR model should be considered. Since the main topic of the present paper is investigating the stability of isotropic and anisotropic exponential inflation, we will leave this issue to our future study.  We hope that our present paper could be useful to studies of cosmological implications of fourth-order gravities. 
\begin{acknowledgments}
 We would  like to thank Prof. S.~D.~Odintsov very much for his useful suggestions.  We would also like to thank the referee very much for comments and suggestions, which are very useful to improve our paper.
\end{acknowledgments}
\appendix
\section{Geometrical quantities of FLRW metric}
In this Appendix, we are going to list the explicit expressions of non-vanishing components of Riemann tensor, $R^{\mu}{ }_{\nu\rho\sigma}$, for the FLRW metric as follows
\begin{align}
R^0{ }_{101}= R^0{ }_{202}=R^0{ }_{303} &= -\frac{e^{2\alpha}} {N^2} \left( \frac{\dot N}{N}\dot\alpha -\ddot\alpha- \dot\alpha^2  \right),\\
R^{1}{ }_{010}=R^2{ }_{020}=R^3{ }_{030}&= \frac{\dot N}{N} \dot\alpha-\ddot\alpha -\dot\alpha^2,\\
R^1{ }_{212}=R^1{ }_{313}=R^2{ }_{121}=R^3{ }_{131}=R^2{ }_{323}=R^3{ }_{232} &=\frac{e^{2\alpha}}{N^2} \dot\alpha^2.
\end{align}
Hence, it is straightforward to obtain the following Ricci tensor, $R_{\mu\nu}\equiv R^\rho{ }_{\mu\rho\nu}$ as follows
\begin{align}
R_{00}&=3 \left( \frac{\dot N}{N} \dot\alpha - \ddot\alpha - \dot\alpha^2  \right),\\
R_{11}=R_{22}=R_{33} &=- \frac{e^{2\alpha}}{N^2} \left( \frac{\dot N}{N} \dot\alpha -\ddot\alpha -3\dot\alpha^2 \right).
\end{align}
Finally, the corresponding Ricci scalar turns out to be
\begin{equation}
R \equiv g^{\mu\nu}R_{\mu\nu} =-\frac{6}{N^2} \left( \frac{\dot N}{N} \dot\alpha   -\ddot\alpha - 2\dot\alpha^2 \right).
\end{equation}
Given these results, we will be able to define the corresponding Gauss-Bonnet term to be
\begin{equation} \label{RGB}
{\cal G} =-\frac{24}{N^5} \dot\alpha^2 \left[ \dot N \dot \alpha - N \left(\ddot\alpha+\dot\alpha^2 \right) \right].
\end{equation}
\section{Geometrical quantities  of Bianchi type I metric}
In this Appendix, we are going to list the explicit expressions non-zero components of Riemann tensor, $R^{\mu}{ }_{\nu\rho\sigma}$, for the Bianchi type I metric as follows
\begin{align}
R^0{ }_{101} &= -\frac{e^{2\alpha-4\sigma}}{N^2} \left[\frac{\dot N}{N} \left(\dot\alpha-2\dot\sigma \right) - \ddot\alpha+2\ddot\sigma- \left(\dot\alpha -2\dot\sigma \right) ^2  \right], \\
 R^0{ }_{202}=R^0{ }_{303} &=- \frac{e^{2\alpha+2\sigma}}{N^2} \left[\frac{\dot N }{N}\left(\dot\alpha+\dot\sigma \right) - \ddot\alpha-\ddot\sigma -\left(\dot\alpha^2+\dot\sigma\right)^2 \right] , \\
R^{1}{ }_{010}  &=\frac{\dot N}{N} \left(\dot\alpha -2\dot\sigma \right) -\ddot\alpha +2\ddot\sigma -\left(\dot\alpha-2\dot\sigma \right)^2 , \\
R^2{ }_{020}=R^3{ }_{030}&= \frac{\dot N}{N} \left(\dot\alpha + \dot\sigma \right) -\ddot\alpha -\ddot\sigma -\left(\dot\alpha+\dot\sigma \right)^2 ,\\
R^1{ }_{212} =R^1{ }_{313} &= \frac{e^{2\alpha+2\sigma}}{N^2} \left(\dot\alpha-2\dot\sigma \right) \left(\dot\alpha+\dot\sigma \right),\\
R^2{ }_{121}=R^3{ }_{131}& =  \frac{e^{2\alpha-4\sigma}}{N^2} \left(\dot\alpha-2\dot\sigma \right) \left(\dot\alpha+\dot\sigma \right),\\
R^2{ }_{323}=R^3{ }_{232} &=\frac{e^{2\alpha+2\sigma}}{N^2} \left(\dot\alpha+\dot\sigma\right) ^2.
\end{align}
Hence, it is straightforward to obtain the following Ricci tensor, $R_{\mu\nu}\equiv R^\rho{ }_{\mu\rho\nu}$, as follows
\begin{align}
R_{00}&=3 \left( \frac{\dot N}{N} \dot\alpha - \ddot\alpha - \dot\alpha^2 -2\dot\sigma^2 \right), \\
R_{11} & =- \frac{e^{2\alpha -4\sigma}}{N^2} \left[ \frac{\dot N}{N} \left(\dot\alpha -2\dot\sigma \right) -\ddot\alpha +2\ddot\sigma -3\dot\alpha^2 +6\dot\alpha\dot\sigma \right],\\
R_{22}=R_{33} & = - \frac{e^{2\alpha +2 \sigma}}{N^2} \left[ \frac{\dot N}{N} \left(\dot\alpha + \dot\sigma \right) -\ddot\alpha - \ddot\sigma -3\dot\alpha^2 - 3\dot\alpha\dot\sigma \right].
\end{align}
Finally, the corresponding Ricci scalar turns out to be
\begin{equation}
R  =-\frac{6}{N^2} \left( \frac{\dot N}{N} \dot\alpha   -\ddot\alpha - 2\dot\alpha^2-\dot\sigma^2 \right).
\end{equation}
Given these results, we will be able to define the corresponding Gauss-Bonnet term to be
\begin{equation} 
{\cal G} =-\frac{24}{N^5}\left( \dot\alpha +\dot\sigma \right) \left\{ \dot N \left(\dot \alpha -2\dot\sigma \right)\left(\dot\alpha+\dot\sigma\right) - N \left[\ddot\alpha \left(\dot\alpha-\dot\sigma \right)-2\ddot\sigma \dot\sigma +\dot\alpha^3 -\dot\alpha\dot\sigma \left(\dot\alpha+2\dot\sigma \right)   \right] \right\}.
\end{equation}
As mentioned above, the corresponding Pontryagin density ${ P_4}$ vanishes identically for the Bianchi type I metric.

\end{document}